# Exploring Query Understanding for Amazon Product Search

Chen Luo, Xianfeng Tang, Hanqing Lu, Yaochen Xie, Hui Liu, Zhenwei Dai, Limeng Cui, Ashutosh
Joshi, Sreyashi Nag, Yang Li, Zhen Li, Rahul Goutam, Jiliang Tang, Haiyang Zhang, Qi He
Query Understanding Team
Amazon Search
Palo Alto, California, USA

## ABSTRACT

Online shopping platforms, such as Amazon, offer services to billions of people worldwide. Unlike web search or other search engines, product search engines have their unique characteristics, primarily featuring short queries which are mostly a combination of product attributes and structured product search space. The uniqueness of product search underscores the crucial importance of the query understanding component. However, there are limited studies focusing on exploring this impact within real-world product search engines. In this work, we aim to bridge this gap by conducting a comprehensive study and sharing our year-long journey investigating how the query understanding service impacts Amazon Product Search. Firstly, we explore how query understanding-based ranking features influence the ranking process. Next, we delve into how the query understanding system contributes to understanding the performance of a ranking model. Building on the insights gained from our study on the evaluation of the query understanding-based ranking model, we propose a query understanding-based multi-task learning framework for ranking. We present our studies and investigations using the real-world system on Amazon Search.

## 1 INTRODUCTION

The proliferation of online shopping platforms, such as Amazon, Tmall, or Alibaba, has revolutionized the retail landscape, providing billions of users worldwide with convenient access to a vast array of products. At the core of these platforms lies the product search engine, a vital component responsible for facilitating seamless product discovery and purchase. This system plays a pivotal

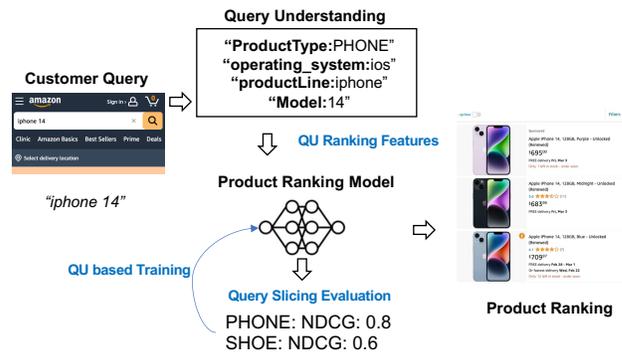

**Figure 1: Investigating the Effectiveness of Query Understanding for Product Search. In this work, we are investing three aspect for query understanding to improve search: QU Ranking Features, Query Segment Evaluation, and QU based Training**

role in capturing customer intent from search queries and presenting a best list of ranked products that align with users' shopping objectives. Efficient product search engines are essential for delivering satisfying user experiences and driving business success for e-commerce platforms. [8, 19]

Different from general web search or other search engines, a product search engine has its unique characteristics. 1. Product search engine keywords mostly consist of product attribute combinations, such as "chair for study" or "red shoes for running." In a recent study conducted on Amazon search, we find that over 80% of search keywords are fewer than 6 words, and more than 90% of the search keywords are attribute combinations rather than natural language questions (e.g., "What is the highest mountain in the world?"). 2. The search space of a product search engine is limited to products only. In contrast, the web search engine's search space includes webpages, videos or other types of items, which are unstructured and consist of heterogeneous data. On the other hand, products have structured data containing information about attributes such as brand, color, size, etc. For example, a shoe product contains structured information about the brand, color, size, and other relevant details.

The unique characteristics of product search make the query understanding component crucially important. The query understanding component serves as an intelligent intermediary, tasked with understanding the explicit attributes mentioned in the search query and deciphering the implicit intent behind users' search queries. By employing sophisticated natural language processing techniques, this component gains insights into users' desires and requirements, enabling it to better grasp the context and nuances of their queries [33]. For example, in [33], the team builds a query parsing model to understand the meaning of every span using a named entity recognition model. Additionally, there are query intent classification techniques extensively used by the query understanding component, classifying the query into different intents that are utilized in product ranking [10]. Understanding the structured attribute information from the query is the first step of a product search engine. The matching and ranking of a product rely on attribute matching [32]. For instance, if a product's attributes match all the attributes from the query keywords, then this product will be a perfect match. On the other hand, if the attribute of the query does not match the attribute of the product, such as a customer searching for "red shoes," and the search result shows blue shoes, where the attribute color does not match between the query and product, it will not be a proper search result.

However, despite these advancements, there are limited studies focusing on exploring the impact of the query understanding component within real-world product search engines. Previous research endeavors [10, 32, 33] have separately explored the potential of query understanding, as well as matching and ranking in product search. Some studies have focused on improving query understanding techniques, enhancing the accuracy of understanding user intent from search queries [19, 20]. Conversely, other research has delved into refining product ranking algorithms to optimize product relevance and improve user satisfaction [32]. These different lines of study do not have a common space that can truly tackle the importance of query understanding for product search engines.

In this paper, we aim to contribute to the field of product search engines by investigating the impact of query understanding on the product search. Specifically, drawing from our year-long journey in real product search studies at Amazon search, we delve into three key aspects: a) Query understanding signals that provide essential features for the ranking model in the product search engine. 2) In the context of product search engines, effectively integrating query understanding signals into the ranking model evaluation process holds significant potential to enhance the overall performance and relevance of search results. 3) Leveraging the insights gained from query understanding, we can augment traditional ranking evaluation methodologies to offer a more comprehensive and context-aware assessment of the ranking model. In the end we propose a new query understanding based multi-task learning framework for training the ranking models. By exploring these crucial facets, our aim is to shed light on the intricate relationship between query understanding and product ranking, ultimately contributing to the advancement of online shopping platforms and the improvement of the search experience for billions of users worldwide.

## 2 PRELIMINARY

### 2.1 Query Understanding

Query Understanding (QU) transforms a search query into an enhanced query that maps to the specific search engine grammar and thus can retrieve more relevant documents. In an e-commerce site, QU can fix spelling mistakes, extract shopping intents and product attributes from search queries, and incorporate behavioral data to facilitate retrieval. Historically, researches have used handcrafted features or rules [7] in conventional QU approaches. However these approaches lack data (and thus have low performance) for tail queries which suffer from limited coverage on tail queries.

Neural nets, trained from scratch [15, 24] or fine-tuned from pre-trained models [9, 11, 14, 33] have also been employed to improve generalization in QU. Either way, the models are trained in a supervised setting requiring a large amount of annotated data. Pre-trained models also need to be adapted for the product search domain. Gu et al. [5], Gururangan et al. [6], Lee et al. [12] enable domain adaptation by adding a second level of in-domain pre-training from a public checkpoint and demonstrate advantages over the open-domain models. Finally, though generative large language models (LLMs) show potential in query understanding, we are not yet aware of any published research that uses LLMs for this purpose. In addition, the slow speed of LLMs make them unattractive for commercial systems that have strict latency requirements.

### 2.2 Product Search

In e-commerce, product ranking connects customers to the huge online inventory, ensuring that the most relevant and popular products are displayed at the top of search results. Improving product ranking directly impacts user experience, conversions rate, and customer satisfaction [4].

Similar to web search ranking, a variety of learning to rank (L2R) methods (e.g., [1, 3, 16, 28, 29]) have been proposed and applied to optimize the product ranking for search engines. Generally, L2R takes various product and query features as input, and predicts scores for products as the ranking result. It then parameterized the projection between input features and output ranking with machine learning models, such as regression, boosting, and neural networks. However, simply applying L2R methods for product search is less optimal, as ranking products is more challenging than ranking web search results due to relatively shorter query keywords and the huge catalog space [31]. To improve the performance of L2R approaches, Trotman et al.[30] propose to first retrieve related products via multiple steps, then re-rank retrieved products so that the item-space is reduced. Long et al.[17] leverage best-selling items for the product ranking. Bi et al.[2] introduce user behaviour data such as engagement and click-through rate to improve the ranking algorithm. Parikh and Sundaresan[25] further consider diversity of the search result to enhance user experience.

Recently, multi-task learning (MTL) has attracted increasing attention in product ranking field. Comparing to conventional single-task approaches, multi-task learning can jointly train multiple tasks/objectives simultaneously. Since these tasks usually share model parameters, they can benefit from each other by exchanging related information in training data and regularize each other during the training process [34]. Because ranking problems naturally has multiple learning objectives, such as the accuracy of ranking scores, diversity of products, and personalized preferences, they can be benefited from MTL. For example, Lu et al.[18] combine perform rating prediction and recommendation explanation in matrix factorization; Wu et al.[31] propose a MTL framework for product ranking that integrates multiple types of engagement signals; and Li and Gaussier[13] propose a cascaded late interaction approach to balance the cost between attenten computation and dense retrieval.

While MTL demonstrates promising results in product ranking area, how to incorporate query understanding into MTL is still under-explored.

## 3 QUERY UNDERSTANDING AT AMAZON

Query Understanding serves as the entry point for Amazon product search engine. In this section, we introduce key components of query understanding at Amazon, including product intention detection, query parsing, sensitive query detection, and conversation and session understanding within Amazon.

### 3.1 Product Intention Detection

Amazon Search supports queries from millions of requests across many marketplaces and in many languages. We must support prediction in each with low latency. One solution would be to build individual models for each market place and each language; however, serving multiple models will dramatically increase our operational cost to run and maintain. As a result, we designed a single, multilingual model which consumes the marketplace (e.g. US, UK, etc) as a signal in prediction. We use a transformer query representation, initialized with the publicly available[1] multilingual DistilBert [27].

Our basic model builds on the transformer architecture. After obtaining the embeddings for each token of the input query, we apply classification on top of the [CLS] output embedding. In this network structure, each marketplace has a different label space, consisting of labels observed for products in that marketplace. We concatenate all the label spaces together, and each input marketplace $m$ masks the corresponding labels for that marketplace.

The problem is a multi-label problem, so we use the binary cross entropy as our loss. Since the output space for the model is large, we use a semi-sparse loss function to speed up the calculation, allowing us to only store nonzero labels:

$$
\begin{aligned}
Loss &= \sum_t f(y_t, \hat{y}_t) \\
&= \sum_t f(0, \hat{y}_t) + \sum_t (f(y_t, \hat{y}_t) - f(0, \hat{y}_t)) \\
&= \sum_t f(0, \hat{y}_t) + \sum_{t, y_t \neq 0} (f(y_t, \hat{y}_t) - f(0, \hat{y}_t)),
\end{aligned}
\tag{1}
$$

where $y_t$ is the ground truth and $\hat{y}_t$ the model prediction for label $t$, and $f()$ is the single-label binary-cross entropy.

We tested the model performance of our multi-marketplace model by comparing it with select single-marketplace models. We use the same evaluation data and trained three representative single marketplace models in US, DE, and JP. The result is shown in Table 1. Performance for US and DE is on par with language-specific models, JP sees a slight performance drop.

### 3.2 Query Parsing

The query parsing module is a NER model aims to recognize the named entities, such as "product type", "brand", "media title", etc, in the search keywords. However, traditional NER model cannot accurately recognize the knowledge intensive entities such as "media titles" since these entites require the model to memorize all the related knowledge in the parameters. For example, if we attempt to use a NER model to annotate the movie title in the query "Yosemite 2015", the term "Yosemite" is more likely to be identified as a national park name rather than a movie title. To correctly recognize it as a movie title, the NER model must have memorized the specific title "Yosemite" during training. Amazon Search develops a retrieval augmented NER (RA-NER) model to address this problem.

---

[1]https://huggingface.co/distilbert-base-multilingual-cased

**Table 1: The recall at precision 85% for world wide model vs. single marketplace model.**

| Marketplace | Worldwide Model | Production |
|---|---|---|
| US | 81.1% | 79.4% |
| DE | 86.6% | 86.4% |
| JP | 80.0% | 82.1% |

RA-NER includes two parts: (1) a retriever to retrieve the most relevant information from an external knowledge database and (2) a NER model that classifies tokens into different entities. Given a query, the retriever will first retrieve the most related information from the external knowledge database. Then, RA-NER will parse the input query along with the retrieved information to the NER model. Finally, the NER model encodes and annotate each token of the input query with a predetermined entity (e.g. Brand, Color, Product Type, Title, etc).

**Retriever:** The retriever aims to retrieve the most relevant information from the external knowledge dataset, where the "relevance" of the information can be measured using semantic or string-level similarity. For example, we can use HNSW (Hierarchical Navigable Small World) to search for information with similar semantics [22], or LSH (locality-sensitive hashing) to retrieve information with high string-level similarity (Jaccard similarity). In the real production system, we use the LSH-based similarity search [21] due to latency and memory constraints.

**NER Model:** The NER model includes an encoder to encode the input tokens into embeddings, and a classification head to classify the token into different entities. Note that classification head only applies to the input query tokens. To improve the classification accuracy, NER model usually uses conditional random field (CRF) to better model the sequence labeling and label annotations [26].

Table 2 shows the experimental results of the RA-NER against the production model. We observe a relatively good improvement in NER for most languages.

| Language | The $F_1$ Score Improvement |
|---|---|
| English | -0.1 |
| German | 0.4 |
| French | 0.1 |
| Italian | 0.5 |
| Japanese | 0.2 |
| Portuguese | 1.3 |

**Table 2: The $F_1$ Scores Improvement for Different Languages**

## 3.3 Conversation and Session Understanding

From the overall search traffic, we identify a noticeable amount of conversation-style queries embodying a wide range of intents. Specifically, the queries are phrased as natural language questions or statements, spanning one or multiple search interaction turns. Unlike the more common keyword-based queries, this format allows users to seek assistance on topics extending beyond simple product searches, or to explore products without prior knowledge of what they are looking for. For example, a customer can ask "What is the mattress return policy?", "Where is my order?", or "What should I prepare for bringing a mid-size puppy home?". To appropriately address these specific types of query, we evolve our keyword-focused query understanding model into the Conversation and Session Understanding (CSU) model. The CSU model enhances the integration of conversational queries with the existing Amazon Search experience by generating three key signals from the conversational queries: question intent classification, context switch detection, and the question-to-keywords (Q2K) rewrite. Utilizing a unified language model, we produce these signals through a sequence-to-sequence approach with a single prompt, training all three tasks concurrently in a supervised fashion using a diverse dataset.

**Question intent classification** The question intent classification process is pivotal in deciphering the diverse intents embedded within natural language questions, which, due to their open-ended nature, may extend beyond mere product inquiries. Recognizing the significance of discerning whether to trigger a product search, we develop the signal that classifies intents into four primary categories including Production search, Helo, General Knowledge, and Sensitive.

(1) Product Search (<P>): This category encompasses queries where the customer is either specifically looking for a product, indicated by a direct question about a known item, or conducting a broader exploration without mentioning a particular product.

(2) Help (<H>): Here, the customer seeks information regarding Amazon's policies, programs, or details about their orders.

(3) General Knowledge (<G>): Queries in this category involve requests for factual information or guidance on products, such as "How far away is the Moon from the Earth?" or "How to interpret COVID-19 test results?"

(4) Sensitive (<S>): This category is reserved for questions that are harmful, unethical, offensive, or of an adult nature.

A search is triggered exclusively in response to queries identified with a Product Search (<P>) intent. We further leverage the task dependencies as a chain-of-thought (CoT) and structure the intent classification as the generation of class tokens together with the other two tasks, thereby streamlining the identification and handling of conversational queries.

**Context switch prediction** The Session and Conversation Understanding (SCU) model is designed to manage multi-turn conversational queries within a shopping session. The model triggers a search page refresh when a new topic is introduced, particularly for general knowledge questions. Therefore, we developed the context switch signal to identify whether the current question maintains the ongoing topic or shifts context. This context switch signal is integrated into a unified response template, serving as a precursor for generating subsequent signals.

**Table 3: Search irrelevance rate (IRR) and sparse result rate (SRR) using original conversational queries and rewritten keywords.**

| Metrics | Single-turn | | Multi-turn | |
|---|---|---|---|---|
| | IRR | SRR | IRR | SRR |
| Vanilla Search | 24.03 | 54.62 | 49.76 | 14.44 |
| With Q2K Rewrite | 5.81 | 0.67 | 24.28 | 3.01 |

**Question-to-keywords rewrite** To seamlessly incorporate the conversational query understanding with current search experience, the CSU model finally generates the question-to-keywords (Q2K) signal. Specifically, the Q2K identifies the shopping intent and rewrites the conversational query, *i.e.*, the question, into search keywords that align with the capabilities of the current search frameworks. The Q2K task is unique in its approach, diverging from conventional text summarization or named entity recognition (NER) techniques, as it adeptly captures shopping intents that may be implicit or indirectly expressed. In scenarios involving multi-turn sessions, the Q2K model plays a crucial role in addressing anaphoric questions, where queries in the latest turn refer back to entities mentioned in earlier interactions. The search is then triggered using the generated keywords, rather than the original conversational query. On real search traffic, we evaluate the search quality improvement on conversational queries brought by the Q2K signal. We quantify the search quality by computing the irrelevance rate (IRR[2]) and sparse result rate (SRR[3]) of search results derived from the original query (question) and rewritten keywords, individually. The scores shown in Table 3 indicates a significant improvement on the search quality introduced by the Q2K intervention.

## 4 QUERY UNDERSTANDING FEATURES

Product ranking models play a critical role in optimizing search results by utilizing behavioral signals such as clicks and purchases. However, these models tend to heavily rely on behavioral features, overlooking the importance of understanding the specific attributes within queries and products. This limitation often leads to challenges in achieving precise matches and ultimately hinders overall performance. For instance, a customer might search for "iPhone" but end up purchasing an "iPhone accessory". Such behavioral signals add certain bias to the ranking model and thus hurt the performance.

We acknowledge the significance of incorporating a query understanding system into the search process. By harnessing advanced named-entity recognition (NER) techniques [33], we can effectively extract and link queries to specific entities, thereby gaining a deeper understanding of customer intent. This approach empowers us to leverage the insights derived from the query understanding system for constructing ranking features and integrating them into the ranking model. The ranking features based on query understanding align query attributes with product attributes, enabling more precise and contextually relevant matches. In this way, our goal is to bridge the gap between customer intent and product offerings, ensuring that users find exactly what they are looking for, thereby enhancing their overall shopping experience on our platform.

---

[2]IRR is computed as the number of irrelevant products in the top-16 search results.
[3]A search keyword has sparse result when there are less than 16 products in search results.

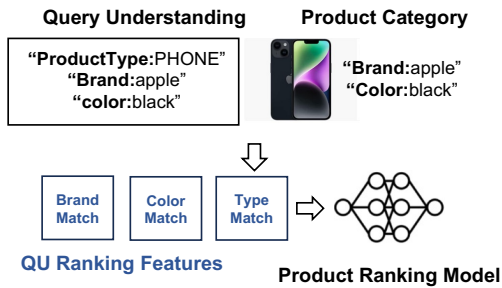

**Figure 2: Query Understanding based Ranking Features**

| Country | W/O QU | W/ QU |
|---|---|---|
| United States | 100% | +0.71% |
| United Kingdom | 100% | +0.82% |
| India | 100% | +0.78% |
| Canada | 100% | +0.59% |
| Japan | 100% | +1.29% |
| Germany | 100% | +0.56% |

**Table 4: Online A/B Testing Results on QU Ranking Feature: Relative Improvement in Amazon Ranking Performance.**

## 4.1 Ranking Features from QU

In this work, we introduce a framework for generating query understanding-based ranking features and assess the impact of these query understanding features on actual ranking performance. At a high level, the construction of ranking features from Query Understanding involves three main components.

- Query Understanding: We require a query understanding module to extract attributes/entities such as brand, color, and product type from customer-input queries.
- Product Catalog: We need to extract the attribute values for each product in the catalog.
- Matching Model: Once we have extracted the information from both the query side and the product side, we can utilize a matching model to compute a feature. The feature could be either boolean or numeric.

Figure 2 depicts the architecture of the framework. The query understanding component parse the query and generates attributes or entities from it. Subsequently, the product catalog data furnishes attributes/entities for each product. The matching model then generates the query understanding (QU) ranking features, such as checking whether the brand in the query matches that in the feature, or if the colors match. There are also query specific ranking features like query specificity, customer intention [19]. These features are used by the ranking models for offline training and online inference.

## 4.2 Empirical Study

We evaluate the impact of QU-based ranking features through product ranking (e.g., [1, 3, 16, 28, 29]) at Amazon Search, aiming to reorder the top-k products based on their relevance to the query intent. In this work, we choose $k = 16$ as top 16 products are usually in the first page in Amazon Search. We first use query understanding to extract attributes from the product search queries [19, 20]. Next, we generate boolean features, such as "is product type a match" and "is brand a match," grounded in the attribute values of both queries and products. These boolean features are referred to as QU ranking features. There are also some attributes that are not product-related, such as query specificity (whether the query is broad, exploring, or looking for specific products) or query statistics, such as purchase rate or click rate. Subsequently, we train two learning-to-rank (LTR) models that are used at Amazon Search: one employs QU ranking features, while the other does not. All other features, configurations, and hyperparameters of these two models remain identical. To compare the two models, we utilize NDCG@16, representing the normalized discounted cumulative gain (NDCG) score for the top 16 products in the search results. We conducted these experiments on different ranking models and different data across six countries: United States, United Kingdom, India, Canada, Japan, and Germany. The relative improvement in NDCG@16 due to QU ranking features is presented in Table 4. On average, we observe an enhancement of 0.79% in NDCG@16, affirming the effectiveness of QU ranking features. This demonstrated that the use of query understanding-based ranking features enables us to enhance ranking performance in the real production system like Amazon Search.

## 5 QUERY UNDERSTANDING BASED MULTI-TASK RANKING

Today, when evaluating the product ranking models, we do so in an aggregated manner. For instance, we compute metrics like NDCG, HERO, etc [32] on the entire test dataset. However, this aggregated approach limits our insights into understanding the model's performance. We lack knowledge of how well the model performs on different types of queries, such as broad queries or spearfishing queries. Additionally, we are unaware of the specific categories where the model may underperform.

To address this limitation, we propose a solution using query understanding (QU) features to gain a deeper understanding of the search models. By incorporating segments over the data, we can analyze the model's performance on various query types and categories, providing valuable insights into its strengths and weaknesses. This approach will enable us to make more informed decisions in optimizing and refining the product ranking models based on their performance in specific scenarios (Fig. 3).

By utilizing this query segment framework, we can acquire the performance of product ranking models for various query segments. Consequently, we can engage in multi-task training for different query segments. This involves training the model on distinct objectives tailored to each query segment, with the goal of enhancing the overall performance of the ranking model.

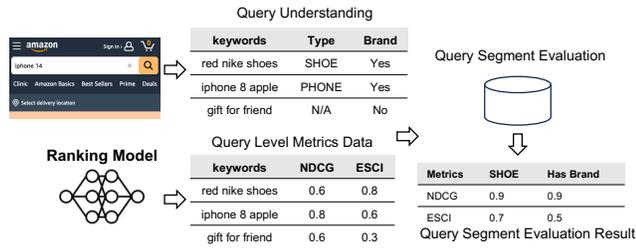

**Figure 3: The framework of using Query Understanding signals to segment the evaluation data and get the detailed evaluation result**

**Table 5: Product Ranking Model Tasks**

| Task name | Definition |
|---|---|
| PurchaseOrClicks | Purchase score and Click score |
| Revenue | The revenue from top K products |
| Relevance | The ESCI of top K products |

Further more, we expand this framework to ranking model training progress. The usual next step after the query segment analysis is to create a separate loss function specifically for that query segments or increase a weight specifically for these low-performing query segments groups. For example, we can set a separate purchase task on the tail queries, i.e queries with low frequency to force the ranking model focus more on the purchase task when the query is a tail query. Similar action can be performed on other query segments for every task but it is exponentially time-consuming to do so on the combinations of all the tasks and all the query segments. Additionally, product ranking model is a multi-task learning model and how to tune the multi-task weights is usually a pain point to find the preferred the pareto front optimal point. More weights on one task, say purchase, means to aim for the optimal point with higher purchase score. Ideally we would want to maximize the revenue while maintaining relevance score by not dropping too much. So it becomes tricky to tune the task weights. Grid Search is a usual way of tuning the task weights, but with a finite number of tasks we have infinite choices of the task weights and that would take time and computation resources. Therefore, it would alleviate this pain point by incorporating the query segment analysis into the model training process and have the model dynamically tune the task weights based the performance of query segments at each step.

## 5.1 High Level Overview

The overarching goal of this algorithm is to incorporate query segment analysis into the model training process by identifying low-performing query segments for each task and subsequently increasing the task weight for those specific segments. In cases of conflicting purchase tasks and relevance tasks, the model is designed to prioritize one task over the other to align with the company's objectives. The determination of low-performing query segments is accomplished through the measurement of task loss for that particular query segment within a support set, enabling the estimation of validation loss. This concept is inspired by the approach introduced in 'Learning to Weight' [23]. There are two primary methods for identifying low-performing query segments: comparing a segment's performance against its own historical data vertically, or evaluating its performance relative to its peers horizontally: 1) Worse performance compared within its history. If the support set loss for that query segments is more than last checkpoint. It means this task is defeated by other tasks and the model sacrifice this task in the favor of other tasks. 2) Low performing compared with other query segments. If the support set loss for that query segments is more than average task loss of other segments, then we should increase the task weight for that query segment because it is falling behind other query segments.

In this work, we design a strategy to update tasks after identifying the low-performing query segments. For each task $t_k$, this strategy is executed at every $N$ steps:

- **Primary Task:**

$$w'_{t,qs} = w_{t,qs} + \alpha \cdot \text{Relu} \left( \text{Loss}_S(i, t, qs) - \text{Loss}_S(i - 1, t, qs) \right)$$
$$\cdot \max \left( 1, \frac{\text{Loss}_S(i, t, qs)}{\text{Loss}_S(i, t, S)} \right)$$

- **Auxiliary Task:**

$$w'_{t,qs} = w_{t,qs} + \text{Relu} \left( \text{Loss}_S(i, t, qs) - \text{Loss}_S(i - 1, t, qs) \right)$$

In above equation, *Loss* denotes the loss function of the ranking model. $N$ controls the frequency of updating task weights per query segments. From our experience, we recommended to have large N, say 500 steps, to show the effect of task weights tuned in the previous checkpoint. $\alpha$ controls how much more penalty we want to have on primary task weights if the support set loss of primary tasks get higher. Here $\alpha > 1$.

In the previous published works, the multi-task weights are tuned together for the whole training dataset. They ignore the possibility that different groups of training data has different learning patterns. For example, high frequency queries is very easy to be learned with high purchase score whereas low frequency queries are struggling with purchase metrics. Our proposed algorithm tunes the task weight in a fine grained fashion that it would focus more weight on purchase metric for low frequency queries than high frequency queries.

The term inside *Relu* is to identify low-performing query segments by comparing itself within its own history. In our product ranking model training at Amazon Search, a common situation is that purchase task validation losses increases during the training while the relevance-related validation losses decreases which shows the competition between purchase task and relevance task and purchase task loses. Adding this term into the algorithm help alleviate such issue by focusing more purchase task. To solve this problem, we compare the support set losses during the training process, if the any of the task support set loss gets higher, we assume that it is losing to other tasks and thus increases its task weight to maintain the balance. For primary tasks, we increases it more than auxiliary tasks. $\alpha$ controls how much more here.

The term inside max is the way of identifying low-performing query segments that is doing worse on primary task. Because the primary task is the main goal of this ranking model, those query segments perform worse on the primary task should focus more weight on primary task and sacrifice the auxiliary tasks. This term simulates the idea that when students in universities need to switch more time from elective courses to required courses when they don't perform as good as others in the required courses.

**Table 6: Evaluation Results**

| Metrics | Base-Model | Manual Search | Dynamic |
|---------|-----------|---------------|---------|
| NDCG@16 | 100.00% | +6.61% | +8.18% |
| IRR@16 | 100.00% | +0.52% | +0.53% |
| HERO@16 | 100.00% | +0.86% | +0.98% |

**Table 7: Query Segments Weight Analysis**

| Query Segments | Purchase | Revenue | Relevance | Brand |
|----------------|----------|---------|-----------|-------|
| MediaLine | 1.65 | 1.903 | 0.4142 | 0.1752 |
| SoftLine | 1.401 | 1.976 | 0.4501 | 0.3054 |
| HardLine | 1.65 | 1.832 | 0.3994 | 0.2089 |
| Consumable | 1.2 | 1.494 | 0.4059 | N/A |
| hasBrand | 1.647 | 1.71 | 0.3811 | 0.1779 |
| hasColor | 1.503 | 1.951 | 0.3861 | 0.2834 |
| PTMatchRatio-High | 1.75 | 1.882 | 0.3953 | 0.1789 |
| PTMatchRatio-Low | 1.519 | 1.81 | 0.4081 | 0.2326 |
| ProductPrice-High | 1.528 | 1.913 | 0.4133 | 0.2333 |
| ProductPrice-Mid | 1.429 | 1.546 | 0.3953 | 0.1911 |
| ProductPrice-Low | 1.941 | 1.538 | 0.3424 | 0.1418 |
| Frequency-High | 1.963 | N/A | N/A | 0.1774 |
| Frequency-Mid | 1.536 | 1.837 | 0.3931 | 0.233 |
| Frequency-Low | 1.63 | 1.716 | 0.3656 | 0.2815 |
| Specificity-High | 1.584 | 2.064 | 0.4837 | 0.2129 |
| Specificity-Mid | 1.729 | 1.807 | 0.3881 | 0.2126 |
| Specificity-Low | 1.748 | 1.744 | 0.3571 | 0.1946 |

We want the $\alpha$ to be not too small to see the effect of weight changes but also don't want it to be too high which would make the task weights fluctuates too much and make training performance unstable. Depending on the average change of loss value per N step. A good selection of $\alpha$ would be $\frac{0.1}{\Delta Loss}$

## 5.2 Empirical Study

*5.2.1 Results Discussion.* We compare our methods with the fixed weight strategy for multi-task ranking model training, referred to as the 'Manual Grid Search' model. A comprehensive comparison of results between the 'Manual Grid Search' model and the query segment-based algorithm is provided in Table 6. Overall, our proposed algorithm achieves superior results compared to the 'Manual Grid Search' model while consuming fewer computational resources and less time. This efficiency is attributed to our algorithm's utilization of approximately 5 $\alpha$ choices, in contrast to the 'Manual Grid Search' model's evaluation of over 100 different task weight parameter selections. Additionally, our algorithm yields greater improvements in revenue-related tasks. Notably, the NDCG@16 model demonstrates a 150 bps improvement in the revenue task when compared to the 'Manual Grid Search' model. In terms of relevance tasks, the query segment-based ranking model showcases comparable performance to the 'Manual Grid Search' model, with an approximately 40 bps improvement.

*5.2.2 Effect of $\alpha$.* As discussed in section 5.1, $\alpha$ controls the weight incremental for primary tasks when the support set loss is decreasing. If $\alpha$ is too small, then primary tasks would need more steps to get enough attention. Additionally, setting $\alpha$ to 0 would turn the ranking model into static weight. If $\alpha$ is too large, the model could fluctuates over the optimal task weights. In our experiment, we observe that with non zero $\alpha$ selection, we sees the primary tasks gets more weight to be learned and the validation loss starts dropping instead of increasing. With larger $\alpha$ value, we see a steeper drop with the first 2000 steps, a better primary task performance but worse relevance task performance.To select an appropriate $\alpha$, this depends on the launching criteria of the ranking model, i.e. the lowest threshold of relevance loss a ranking model needs to have. In the experiments we run, we chose the $\alpha$ equal to 10 which meets the bar of relevance score and a very good primary tasks performance.

*5.2.3 Query Segments Weight Analysis.* The overall query segment weights, using purchase task as an example, increases quickly in the beginning stage and then slows down, reaching the optimal weight in small steps. Other tasks follow the same pattern except that the auxiliary task weights would be decreasing to find the optimal weight since the ranking model would like to focus more on the purchase tasks. See Table 7 for the final query segment weights. Use query specificity as an example. The low query specificity has lower revenue task weight than the revenue task weight of high spedicity and mid specificity. Because if the query is low specificity, it means the customers are very aware of the products they would like to buy and make orders once they find the desired products. Usually, the low specificity queries would bring the most revenue compared to mid specificity and high specificity.

## 6 CONCLUSION

The evolution of online shopping platforms, exemplified by giants like Amazon, has revolutionized the global marketplace, offering unparalleled services to billions worldwide. Unlike traditional web searches, product search engines possess distinctive attributes, predominantly characterized by succinct queries composed of product attributes within a structured search space. The significance of the query understanding component in product search cannot be overstated, yet empirical investigations into its impact within real-world product search engines remain limited.

In this study, we endeavored to address this gap by embarking on a comprehensive exploration of the influence of query understanding on Amazon Product Search. Over the course of a year, our journey led us to dissect the intricate interplay between query understanding and product ranking. Our findings underscore the indispensable role of query understanding in optimizing product search engines, paving the way for enhanced search experiences and heightened user satisfaction in the ever-expanding landscape of online commerce.

# REFERENCES

[1] Leif Azzopardi and Guido Zuccon. 2016. Advances in formal models of search and search behaviour. In *Proceedings of the 2016 ACM International Conference on the Theory of Information Retrieval*. 1–4.

[2] Keping Bi, Choon Hui Teo, Yesh Dattatreya, Vijai Mohan, and W Bruce Croft. 2019. Leverage implicit feedback for context-aware product search. *arXiv preprint arXiv:1909.02065* (2019).

[3] Olivier Chapelle and Yi Chang. 2011. Yahoo! learning to rank challenge overview. In *Proceedings of the learning to rank challenge*. PMLR, 1–24.

[4] Nurendra Choudhary, Nikhil Rao, Sumeet Katariya, Karthik Subbian, and Chandan K Reddy. 2022. ANTHEM: Attentive hyperbolic entity model for product search. In *Proceedings of the Fifteenth ACM International Conference on Web Search and Data Mining*. 161–171.

[5] Yu Gu, Robert Tinn, Hao Cheng, Michael Lucas, Naoto Usuyama, Xiaodong Liu, Tristan Naumann, Jianfeng Gao, and Hoifung Poon. 2021. Domain-specific language model pretraining for biomedical natural language processing. *ACM Transactions on Computing for Healthcare (HEALTH)* 3, 1 (2021), 1–23.

[6] Suchin Gururangan, Ana Marasović, Swabha Swayamdipta, Kyle Lo, Iz Beltagy, Doug Downey, and Noah A Smith. 2020. Don't Stop Pretraining: Adapt Language Models to Domains and Tasks. *arXiv preprint arXiv:2004.10964* (2020).

[7] Peter V Henstock, Daniel J Pack, Young-Suk Lee, and Clifford J Weinstein. 2001. Toward an improved concept-based information retrieval system. In *Proceedings of the 24th annual international ACM SIGIR conference on Research and development in information retrieval*. 384–385.

[8] Sharon Hirsch, Ido Guy, Alexander Nus, Arnon Dagan, and Oren Kurland. 2020. Query reformulation in E-commerce search. In *Proceedings of the 43rd International ACM SIGIR Conference on Research and Development in Information Retrieval*. 1319–1328.

[9] Haoming Jiang, Danqing Zhang, Tianyu Cao, Bing Yin, and Tuo Zhao. 2021. Named Entity Recognition with Small Strongly Labeled and Large Weakly Labeled Data. *arXiv preprint arXiv:2106.08977* (2021).

[10] Weize Kong, Rui Li, Jie Luo, Aston Zhang, Yi Chang, and James Allan. 2015. Predicting search intent based on pre-search context. In *Proceedings of the 38th International ACM SIGIR Conference on Research and Development in Information Retrieval*. 503–512.

[11] Mukul Kumar, Youna Hu, Will Headden, Rahul Goutam, Heran Lin, and Bing Yin. 2019. Shareable Representations for Search Query Understanding. *arXiv preprint arXiv:2001.04345* (2019).

[12] Jinhyuk Lee, Wonjin Yoon, Sungdong Kim, Donghyeon Kim, Sunkyu Kim, Chan Ho So, and Jaewoo Kang. 2020. BioBERT: a pre-trained biomedical language representation model for biomedical text mining. *Bioinformatics* 36, 4 (2020), 1234–1240.

[13] Minghan Li and Eric Gaussier. 2022. Bert-based dense intra-ranking and contextualized late interaction via multi-task learning for long document retrieval. In *Proceedings of the 45th International ACM SIGIR Conference on Research and Development in Information Retrieval*. 2347–2352.

[14] Zheng Li, Danqing Zhang, Tianyu Cao, Ying Wei, Yiwei Song, and Bing Yin. 2021. Metats: Meta teacher-student network for multilingual sequence labeling with minimal supervision. In *Proceedings of the 2021 Conference on Empirical Methods in Natural Language Processing*. 3183–3196.

[15] Heran Lin, Pengcheng Xiong, Danqing Zhang, Fan Yang, Ryoichi Kato, Mukul Kumar, William Headden, and Bing Yin. 2020. Light Feed-Forward Networks for Shard Selection in Large-scale Product Search. (2020).

[16] Tie-Yan Liu et al. 2009. Learning to rank for information retrieval. *Foundations and Trends® in Information Retrieval* 3, 3 (2009), 225–331.

[17] Bo Long, Jiang Bian, Anlei Dong, and Yi Chang. 2012. Enhancing product search by best-selling prediction in e-commerce. In *Proceedings of the 21st ACM international conference on Information and knowledge management*. 2479–2482.

[18] Yichao Lu, Ruihai Dong, and Barry Smyth. 2018. Why I like it: multi-task learning for recommendation and explanation. In *Proceedings of the 12th ACM Conference on Recommender Systems*. 4–12.

[19] Chen Luo, Rahul Goutam, Haiyang Zhang, Chao Zhang, Yangqiu Song, and Bing Yin. 2023. Implicit Query Parsing at Amazon Product Search. In *Proceedings of the 46th International ACM SIGIR Conference on Research and Development in Information Retrieval*. 3380–3384.

[20] Chen Luo, William Headden, Neela Avudaiappan, Haoming Jiang, Tianyu Cao, Qingyu Yin, Yifan Gao, Zheng Li, Rahul Goutam, Haiyang Zhang, et al. 2022. Query attribute recommendation at Amazon Search. In *Proceedings of the 16th ACM Conference on Recommender Systems*. 506–508.

[21] Chen Luo, Vihan Lakshman, Anshumali Shrivastava, Tianyu Cao, Sreyashi Nag, Rahul Goutam, Hanqing Lu, Yiwei Song, and Bing Yin. 2022. Rose: Robust caches for amazon product search. In *Companion Proceedings of the Web Conference 2022*. 89–93.

[22] Yu A Malkov and Dmitry A Yashunin. 2018. Efficient and robust approximate nearest neighbor search using hierarchical navigable small world graphs. *IEEE transactions on pattern analysis and machine intelligence* 42, 4 (2018), 824–836.

[23] Yuren Mao, Zekai Wang, Weiwei Liu, Xuemin Lin, and Pengtao Xie. 2022. MetaWeighting: Learning to Weight Tasks in Multi-Task Learning. In *Findings of the Association for Computational Linguistics: ACL 2022*. Association for Computational Linguistics, Dublin, Ireland, 3436–3448. https://doi.org/10.18653/v1/2022.findings-acl.271

[24] Priyanka Nigam, Yiwei Song, Vijai Mohan, Vihan Lakshman, Weitian Ding, Ankit Shingavi, Choon Hui Teo, Hao Gu, and Bing Yin. 2019. Semantic product search. In *Proceedings of the 25th ACM SIGKDD International Conference on Knowledge Discovery & Data Mining*. 2876–2885.

[25] Nish Parikh and Neel Sundaresan. 2011. Beyond relevance in marketplace search. In *Proceedings of the 20th ACM international conference on Information and knowledge management*. 2109–2112.

[26] Nita Patil, Ajay Patil, and BV Pawar. 2020. Named entity recognition using conditional random fields. *Procedia Computer Science* 167 (2020), 1181–1188.

[27] Victor Sanh, Lysandre Debut, Julien Chaumond, and Thomas Wolf. 2019. DistilBERT, a distilled version of BERT: smaller, faster, cheaper and lighter. In *NeurIPS EMC² Workshop*. arXiv:1910.01108 http://arxiv.org/abs/1910.01108

[28] Daria Sorokina and Erick Cantu-Paz. 2016. Amazon search: The joy of ranking products. In *Proceedings of the 39th International ACM SIGIR conference on Research and Development in Information Retrieval*. 459–460.

[29] Niek Tax, Sander Bockting, and Djoerd Hiemstra. 2015. A cross-benchmark comparison of 87 learning to rank methods. *Information processing & management* 51, 6 (2015), 757–772.

[30] Andrew Trotman, Jon Degenhardt, and Surya Kallumadi. 2017. The Architecture of eBay Search.. In *eCOM@ SIGIR*.

[31] Xuyang Wu, Alessandro Magnani, Suthee Chaidaroon, Ajit Puthenputhussery, Ciya Liao, and Yi Fang. 2022. A Multi-task Learning Framework for Product Ranking with BERT. In *Proceedings of the ACM Web Conference 2022*. 493–501.

[32] Tao Yang, Chen Luo, Hanqing Lu, Parth Gupta, Bing Yin, and Qingyao Ai. 2022. Can clicks be both labels and features? Unbiased Behavior Feature Collection and Uncertainty-aware Learning to Rank. In *Proceedings of the 45th International ACM SIGIR Conference on Research and Development in Information Retrieval*. 6–17.

[33] Danqing Zhang, Zheng Li, Tianyu Cao, Chen Luo, Tony Wu, Hanqing Lu, Yiwei Song, Bing Yin, Tuo Zhao, and Qiang Yang. 2021. Queaco: Borrowing treasures from weakly-labeled behavior data for query attribute value extraction. In *Proceedings of the 30th ACM International Conference on Information & Knowledge Management*. 4362–4372.

[34] Yu Zhang and Qiang Yang. 2018. An overview of multi-task learning. *National Science Review* 5, 1 (2018), 30–43.